\documentclass[aps,pra,showpacs,preprintnumbers,amsmath,amssymb,twocolumn]{revtex4}

\usepackage{graphicx}
\usepackage{dcolumn}
\usepackage{bbm}
\usepackage{subfigure}

\newcommand{\be}{\begin{equation}}
\newcommand{\ee}{\end{equation}}
\newcommand{\ba}{\begin{array}}
\newcommand{\ea}{\end{array}}
\newcommand{\bqa}{\begin{eqnarray}}
\newcommand{\eqa}{\end{eqnarray}}

\begin{document}

\title{Robust tripartite-to-bipartite entanglement localization by weak measurements and reversal}
\author{Chunmei Yao$^{1,2\ast}$}
\author{Zhi-Hao Ma$^{3}$}
\author{Zhi-Hua Chen$^{4}$}
\author{Alessio Serafini$^{2}$}
\affiliation{$^1$Department of Physics and Electronic Science, Hunan University of Arts and Science, Changde, 415000, China\\
$^2$Department of Physics and Astronomy, University College London, Gower St., WC1E 6BT London, United Kingdom,\\
$^3$ Department of Mathematics, Shanghai Jiaotong University, Shanghai, 200240,  China \\
$^4$Department of Science, Zhijiang College, Zhejiang University of Technology, Hangzhou, 310024,  China}
\thanks{yyccmei@sina.com}

\date{\today}

\begin{abstract}
We propose a robust and efficient approach for tripartite-to-bipartite entanglement localization. By using weak measurements and quantum measurement reversal, an almost maximal entangled state shared by two parties can be generated with the assistance of the third party by local quantum operations and classical communication from a W-like state. We show that this approach works well in the presence of losses and phase diffusion. Our method provides an active way to fight against decoherence, and might help for quantum communication and distributed quantum computation.
\end{abstract}

\pacs{03.67.-a,03.67.Mn,03.65.Yz,03.65.Ud}

\maketitle
The creation and distribution of bipartite entanglement between remote parties is an important issue in order to realize quantum information processing, especially for quantum key distribution \cite{A.K.}, quantum teleportation \cite{C.H.} and one-way quantum computation \cite{R}. In situations of practical interest, most of these scenarios involve many parties, and the specific pairs of systems which need to share bipartite entanglement in two-parties protocols are not known when the entangled resources are created and distributed among all the parties. It is hence interesting to consider efficient ways to extract bipartite quantum correlations from multipartite entangled states.

One of the standard paradigms to localize bipartite entanglement is related to the notion of entanglement of assistance as defined in
Refs.~\cite{D.P.}. It quantifies the entanglement which can be created by reducing a multipartite entangled state to an entangled state with fewer parties (e.g., bipartite) via measurements. Such a production of entanglement, also called ``assisted entanglement'', is a special case of the localizable entanglement \cite{F.}.

These theoretical constructions are valid for ideal, isolated systems. Since no system can be completely isolated, however, it is also important to understand and optimise techniques to achieve entanglement localization in the face of decoherence and noise. Several techniques have been proposed for keeping quantum states away from environment-induced decoherence, including dynamical decoupling \cite{L}, decoherence-free subspaces \cite{D.A.}, and quantum error correction \cite{J.}. Weak quantum measurements, which can be reversed, provide yet another way to suppress decoherence \cite{N.K} and entanglement decay \cite{Qing}, which was originally considered in the context of quantum error correction \cite{M.K}, and recently demonstrated to suppress amplitude-damping decoherence and protect quantum entanglement \cite{Yong}.

Multipartite entanglement, which leads to richer correlations than bipartite entanglement, is known to be very fragile and to display subtle dynamical features with regard to decoherence \cite{A.A}, especially when the distribution of parts of an entangled multipartite state between several remote recipients is concerned \cite{H.J.}.

Thus, for a single multipartite state, it is important to know the optimal ways in which an Einstein-Podolsky-Rosen (EPR) pair can be obtained between two parties without having to introduce more entanglement into the system. In this paper, after having introduced
the notion of entanglement concentration through weak measurement and reversal, and having illustrated it in a simple case with no dynamics
(Section \ref{entloc}), we focus on the performance of this weak measurement based schemes when confronted with noisy channels (Section \ref{deco}), and also provide the reader with a proposed experimental implementation to test these methods in the lab (Section \ref{exp}).

\section{Entanglement localization by weak measurements}\label{entloc}

Let us now describe our schemes. A sender, called Port, prepared the following W-like state
\begin{equation}
\label{eq:1}
\vert W \rangle_{123}=(a_1\vert 100 \rangle+a_2\vert 010 \rangle+a_3\vert 001 \rangle)_{123},
\end{equation}
where $a_1=a_2=\frac{1}{2},a_3=\frac{1}{\sqrt{2}}$.

Firstly, Port transmits qubits 1 and 2 to two remote parties, named as Alice and Bob, respectively. To better illustrate our future strategy, we consider the very basic problem of deciding if this tripartite entangled pure state can be converted, with a nonzero success probability, into an EPR pair shared between Alice and Bob, with the assistance of Port by local operations and classical communication  (LOCC). The reduced density matrix by tracing over qubit 3 is given by $\rho_{12} = Tr_3(|W\rangle_{123}\langle W|)$.  The concurrence \cite{W.K.} $C_{12}$, which quantifies the entanglement between qubits 1 and 2, is 0.5. A related measure for the multipartite $W$-like state $|W\rangle$ is the concurrence of assistance (COA) \cite{G.G} $C_{12}^a$, which is defined as $C_{12}^a (W_{123}):= \max\sum\limits_i(p_i C(| \phi_{12}^i \rangle))$, where the maximization is taken over all decompositions $\sum\limits_i p_i |\phi_{12}^i\rangle\langle\phi_{12}^i |=tr_3(|W\rangle_{123}\langle W|)$ with respect to Port's ``assistance''. It is employed to quantify the maximal average entanglement between Alice and Bob with the assistance of Port by LOCC, and is also equal to 0.5. Just as in Ref.~[17], a deterministic transformation $|W\rangle_{123}\rightarrow |\phi_{12}\rangle$ is possible if the relation $C_{12}^a (|W\rangle_{123})\geq C(|\phi^{12}\rangle)$ is satisfied, and the COA is a proven to be an entanglement monotone, which means that its value cannot increase on average under LOCC operations.

In addition, Port directly makes a projective measurement in the computational basis $\{|0\rangle, |1\rangle\}$ on his qubit 3, if $|0\rangle_3$ is obtained, an EPR pair shared by Alice and Bob can be achieved with $\frac{1}{2}$ probability, otherwise if $|1\rangle_3$ is obtained, the state of qubits 1 and 2 is separable. In fact, it is easy to see that the equality $C_{12}^a=\frac{1}{2}$, $p_0 C_{12}^0+p_1 C_{12}^1=\frac{1}{2}+0=\frac{1}{2}$ is satisfied, such that an EPR pair can be obtained by a projective measurement with a certain success probability. If weak quantum measurement are introduced, how can we get an EPR pair starting from a $W$-like state? 

Next, let us analyze the change of the entanglement shared by Alice and Bob due to a weak measurement and its reversal. Before qubits 1 and 2 are sent to Alice and Bob, Port makes a dicothomic weak measurement $M_w^i(i=1,2)$  on qubit 1 and 2, respectively, which, conditional to a postselected outcome, can be written as the following non-unitary quantum operation acting by congruence on density matrices in the computational basis:
\begin{equation}
\label{eq:2}
M_{w}^1\otimes M_{w}^2=\left(
\begin{array}{cc}
1 & 0  \\
0 & \sqrt{1-p_1} \\%
\end{array}%
\right)\otimes
\left(
\begin{array}{cc}
1 & 0  \\
0 & \sqrt{1-p_2} \\%
\end{array}%
\right),
\end{equation}
where $p_1$ and $p_2$ are weak measurement strengths. We assume that there is no signal detected [9], 
such that the measurement will partially collapse the state towards $|0\rangle_s$ with a probability $p_w = 1-\frac{p_1}{4}-\frac{p_2}{4}$.
This kind of weak measurement is usually referred to as a `null weak
measurement', and should not be confused with
a `weak value', which corresponds to a conditioned averaging of weak measurements.

After receiving the qubits 1 and 2, Alice and Bob carry out an optimal quantum measurement reversal
operation $M_r^i (i=1,2)$ on their qubit, respectively (this procedure will be dubbed ``distributed strategy'', in that the original measurements and their reversal take place at different nodes). The two-qubit reversal operation is represented by
\begin{equation}
\label{eq:3}
M_{r}^1\otimes M_{r}^2=\left(
\begin{array}{cc}
\sqrt{1-q_1} & 0  \\
0 & 1 \\%
\end{array}%
\right)\otimes
\left(
\begin{array}{cc}
\sqrt{1-q_2} & 0  \\
0 & 1 \\%
\end{array}%
\right),
\end{equation}
where $q_1$ and $q_2$ are reversal measurement strengths. The reversal operation is achieved with a probability $p_r=\frac{(1-q_1)(2-p_2-q_2)+(1-q_2)(2-p_1-q_1)}{4 p_w}$. 
Here, the optimal reversing measurement strength, 
yielding maximum concurrence $C_{12}^w$ [as shown in Eq. (4)], is $q_i\rightarrow 1$, $(i=1,2)$.
Clearly, $q_i=1$ for $i=1,2$ correspond to strong (projective) measurements. 
This may seem counterintuitive, since one could expect local projective measurements, performed directly on the relevant degrees of freedom, to destroy any 
quantum correlations. Quite interestingly, this is not the case here because of how the reversed measurement is chosen with respect to the output state of the first measurement. 
However, it should be noted that the maximal entanglement comes at the expense of the success rate which, because of the form of the initial $W$-state, falls to zero for $q_i\rightarrow 1$.
There is hence a trade-off between success probability and degree of entanglement, such that the desirable value of $q_i$ might change depending on the practical needs involved.
In the following, we shall set $q_i =0.99$.
Notice also that, as one should expect, projective measurements at the first stage of the protocol ($p_i=1$ for $i=1,2$) immediately destroy any correlations and result in no entanglement at all.

After the sequence of weak measurement and its reversal, the concurrence $C_{12}^{w}$ (quantifying the entanglement shared by Alice and Bob) is
\begin{equation}
\label{eq:4}
C_{12}^{w}=\frac{X_{r}}{Y_{r}}
\end{equation}
with the success probability $p_{d}=p_{w}p_{r}$.
Above, $X_{r}=\frac{1}{2}\sqrt{X_{1}}, X_{1}=(1-p_1)(1-p_2)(1-q_1)(1-q_2)$ and $Y_{r}= \frac{1}{4}(1-p_{1})(1-q_{2})+(1-q_{1})[\frac{1}{2}(1-q_{2})
+\frac{1}{4}(1-p_{2})]$.

Alternately, before or after qubits 1 and 2 are sent to Alice and Bob, Port can carry out a weak measurement and its optimal reversing measurement on his qubit 3 (which will be called ``fully local'' strategy, in that measurements and reversal take place at the same site), the corresponding Kraus operations of one qubit could be written as
\begin{equation}
\label{eq:5}
M_{w}^3=\left(
\begin{array}{cc}
1 & 0  \\
0 & \sqrt{1-p_3} \\%
\end{array}%
\right),M_{r}^3=\left(
\begin{array}{cc}
\sqrt{1-q_3} & 0  \\
0 & 1 \\%
\end{array}%
\right),
\end{equation}
where $p_3$ and $q_3$ are the weak measurement and reversal measurement strengths, respectively, and we assume that there is no signal detected  \cite{N.K}. The weak measurement and its reversal will be implemented with probabilities $p_{w'}=1-\frac{p_3}{2}$ , and $p_{r'}=1-\frac{p_3+q_3}{2}$, respectively. Then, the concurrence $C_{12}^{w'}$ can be written as
\begin{equation}
\label{eq:6}
C_{12}^{w'}=\frac{(1-q_3)}{(1-p_3)+(1-q_3)}
\end{equation}
with success probability $p_{f}=p_{w'}p_{r'}$.

As shown in Fig.~1, weak measurements and their reversal can indeed be useful for probabilistically increasing the concurrence $C_{12}^w$ or $C_{12}^{w'}$, and these effects are mainly decided by the strengths $p_i$ and $q_i(i=1,2,3)$. Note that as long as one chooses the strengths $p_i$ and $q_i (i=1,2,3)$ properly, an almost EPR pair can be faithful obtained from the originally W-like state with a certain probability by using of weak measurement and its reversal, regardless of which strategy (1 or 2) is used. In Fig. 1a, for a large $q_i (i=1,2)$, i.e. $q_{i}\rightarrow1$,  the stronger the strength $p_{i} (i=1,2$) the smaller the concurrence $C_{12}^w$, as opposed to Fig.~1b, where the stronger the strength $p_3$ the bigger the concurrence $C_{12}^{w'}$, especially for a large $p_3$. If $p_3\rightarrow1$,  even for an arbitrary value of $q_3$, an EPR pair will be approximated with arbitrary precision with a certain finite probability.

It is interesting to notice that if the concurrence $C_{12}^w$ (or $C_{12}^{w'}$) increases, the corresponding concurrence $C_{13}$ (or $C_{23}$) will decrease via distributed (or fully local) weak measurement and its reversal. That is to say, the entanglement can be effectively transferred among two parties pair of a
$W$-like state by use of weak measurements and their optimal reversal.

\begin{figure}
\subfigure[]
{\begin{minipage}[b]{0.2\textwidth}
\includegraphics[width=1\textwidth]{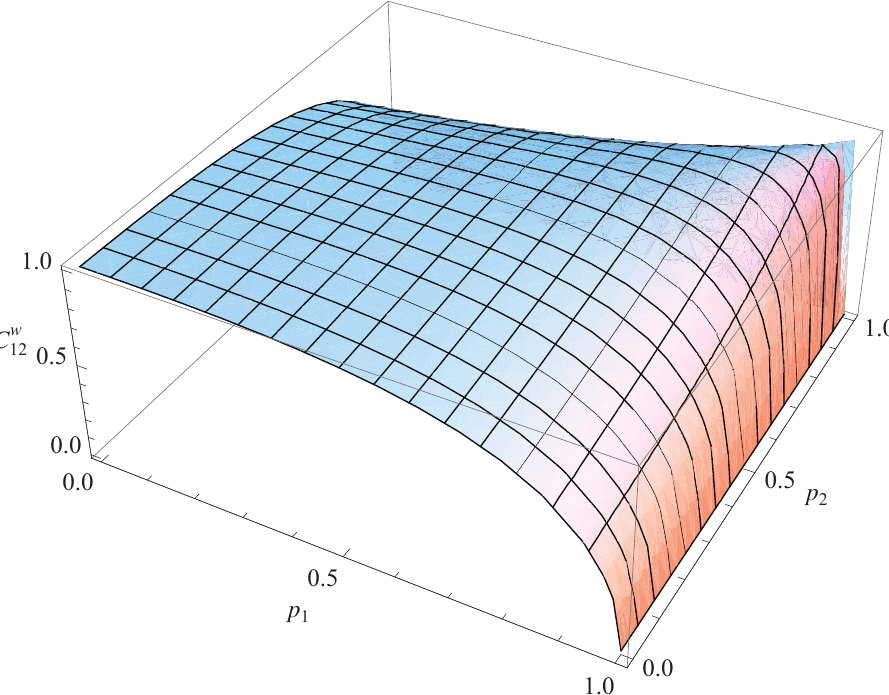}
\end{minipage}}
\subfigure[]
{\begin{minipage}[b]{0.2\textwidth}
\includegraphics[width=1\textwidth]{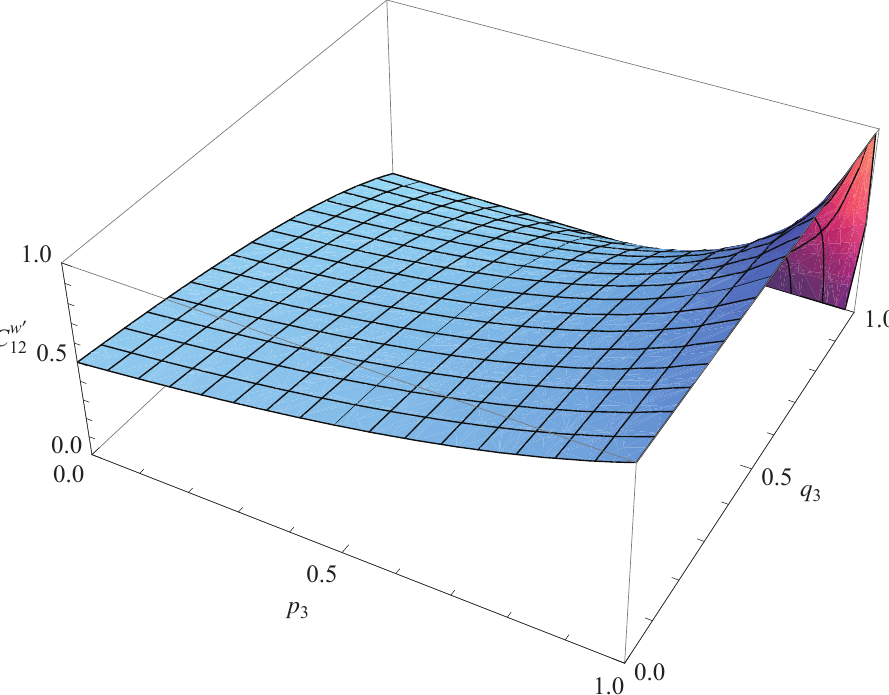}
\end{minipage}}
\caption{Concurrence as a function of weak measurement and its reversal's strengths.
In (a),  the  concurrence $C_{12}^{w}$ under the distributed strategy ($q_1$ $=$ $q_2$ $=0.99$) is reported; in (b), the  concurrence $C_{12}^{w'}$ under a fully local strategy is reported.}
\label{fig:Fig 1}
\end{figure}

\section{Protection of entanglement localization against decoherence}
\label{deco}

In practice, as already remarked, quantum states are unavoidably disturbed by environmental noise during the transmission process. It is thus natural to ask if this weak measurement approach can be applied to taming typical types of quantum noise, such as amplitude-damping (AD) noise and depolarizing (DP) noise. For simplicity, we first deal with AD noise below and then switch to consider DP noise.

The environment (E) is initially a vacuum with zero temperature,  AD noise corresponds to the following map [18]

\begin{equation}
\label{eq:7}
\begin{split}
\vert 0 \rangle_{S}\vert 0 \rangle_{E}\mapsto \vert 0 \rangle_{S}\vert 0 \rangle_{E},\\
\vert 1 \rangle_{S}\vert 0 \rangle_{E}\mapsto \sqrt{1-D}\vert 1 \rangle_{S}\vert 0 \rangle_{E}+\sqrt{D}\vert 0 \rangle_{S}\vert 1 \rangle_{E},
\end{split}
\end{equation}
where $D\in[0,1]$ is the probability of losing the system excitation into the environment.  Suppose Port prepares a $W$-like state in Eq.~(1), and now distributes qubit 1 to Alice and qubit 2 to Bob, respectively, through quantum channels with decoherence $D_1$ and $D_2$.
The W-like pure state inevitably evolves into a mixed state under the disturbance of noise. The effect of decoherence of $D_1$ and $D_2$ on the initial state can be investigated by evaluating the concurrence $C_{12}^{D}$, which is calculated to be
\begin{equation}
\label{eq:8}
C_{12}^{D}=\frac{1}{2}\sqrt{(1-D_{1})(1-D_{2})}.
\end{equation}

From Eq.~(8), it is clear that the stronger the decoherence $D_i (i=1,2)$,  the smaller the  concurrence $C_{12}^{D}$, as depicted in Fig. 2a. In particular, as $D_{i}\rightarrow1$, $C_{12}^{D}\rightarrow0$,  meaning that the two-qubit system becomes fully separable.

In the following section, we describe how to efficiently tame decoherence by using reversed weak measurements. Let us recall that, under what we called ``distributed'' strategy, before qubits 1 and 2 undergo decoherence, Port makes a weak measurement $M_{w}^{i} ( i=1,2)$,  which is written as in Eq. (2), on qubit 1 and 2 respectively. After the noisy quantum channel,  Alice and Bob will carry out corresponding optimal reversal operation $M_{r}^{i} (i=1,2)$ written as in Eq. (3) on their qubits. As the end of the sequence of weak measurement, decoherence and reversal, the concurrence $C_{12}^{r}$ of the two qubits state will be
\begin{equation}
\label{eq:9}
C_{12}^{r}=\frac{X_{r}}{Y_{r}},
\end{equation}
where $X_{r}=\sqrt{X_{1}}, X_{1}=(1-D_{1})(1-D_{2})(1-p_{1})(1-p_{2})(1-q_{1})(1-q_{2})$ and $Y_{r}= \frac{1}{2}(1-p_{1})(1-D_{1}q_{1})(1-q_{2})+(1-q_{1})[(1-q_{2})
+\frac{1}{2}(1-p_{2})(1-D_{2}q_{2})]$

The success probability of the whole event is
\begin{equation}
\label{eq:10}
\begin{split}
P_{suc}=\frac{(1-q_1 )(1-p_2 )(1-q_2 D_2)}{4}+\\
\frac{(1-q_2 )(1-p_1 )(1-q_1 D_1 )}{4}+
\frac{(1-q_1)(1-q_2)}{2}\, .
\end{split}
\end{equation}

We can draw two important conclusions from the result in equation (9), also depicted in Fig.~2b. First, $C_{12}^{r}$ is larger than $C_{12}^{D}$, which means that a weak measurement and its optimal reversal indeed work well in taming decoherence and protecting entanglement although, obviously, at the expense of a finite probability of losing the entanglement altogether. In particular, the W-like state manifests different properties from bipartite entangled states via the distributed method, i.e.~it is more effective in suppressing decoherence and protecting bipartite entanglement. In Ref.[12], a bipartite entangled state can be recovered with a certain probability by the sequence of weak measurements, but the entanglement of the recovered state is smaller than that of the initial state in most cases. Here, $C_{12}^r$ is always larger than $C_{12}^D$, even larger than $C_{12}$ all the time, as the sequence weak measurements can efficiently transfer entanglement among different pairs of a W-like state. Also, it is possible to achieve $C_{12}^{r}\rightarrow1$ (that is, an almost perfect EPR pair shared by Alice and Bob can be generated) with a certain success probability, given by Eq.~(10), provided that the strength of  weak measurement $p_{i}(i = 1, 2)$ is small, and that of the corresponding optimal reversing measurement $q_{i}(i = 1, 2)$ is sufficiently large $(q_{i} \rightarrow1)$, even for very strong decoherence rates $D_{i} (i=1,2)$. From Eq.~(10), it is clear that the larger the entanglement $C_{12}^r$, the less the success probability.

Next, let us discuss the effectiveness of the fully local method for AD noise. Firstly, Port makes a weak measurement $M_w^3$ on qubit 3. After qubits 1 and 2 are transmitted to Alice and Bob, respectively, through the decohering quantum channel, Port carries out the reversal operation $M_{r}^{3}$ on qubit 3. The concurrence $C_{12}^{r'}$ is then
\begin{equation}
\label{eq:11}
C_{12}^{r'}=\frac{X_{r}}{Y_{r}},
\end{equation}
where $X_{r}=\sqrt{(1-D_{1})(1-D_{2})}(1-q_3)$ and $Y_{r}= (1-p_{3})+(1-q_{3})$, with a success probability $1-\frac{p_3+q_3}{2}$

In Fig.~2, we show how $C_{12}^{D}-D_i(i=1,2)$, $C_{12}^{r}-p_i(i=1,2)$ and $C_{12}^{r'}-p_3-D_1(D_2=D_1)$ evolve under different decoherence, weak measurements and optimal reversals. It is clear that decoherence affecting the entanglement between two qubits independently and at different magnitudes can be circumvented by employing reversed weak measurement. That is to say, an almost perfect EPR pair can be faithfully generated from the originally $W$-like state over AD noise channels with a certain success probability via local weak measurements and reversal at the very same site, provided that appropriate strengths $p_i$ and $q_i(i=1,2,3)$ are adopted. Also, quite importantly, it can be seen that the distributed method outperforms the fully local method in the face of AD noise.

\begin{figure}
\subfigure[]
{ \begin{minipage}[b]{0.2\textwidth}
\includegraphics[width=1\textwidth]{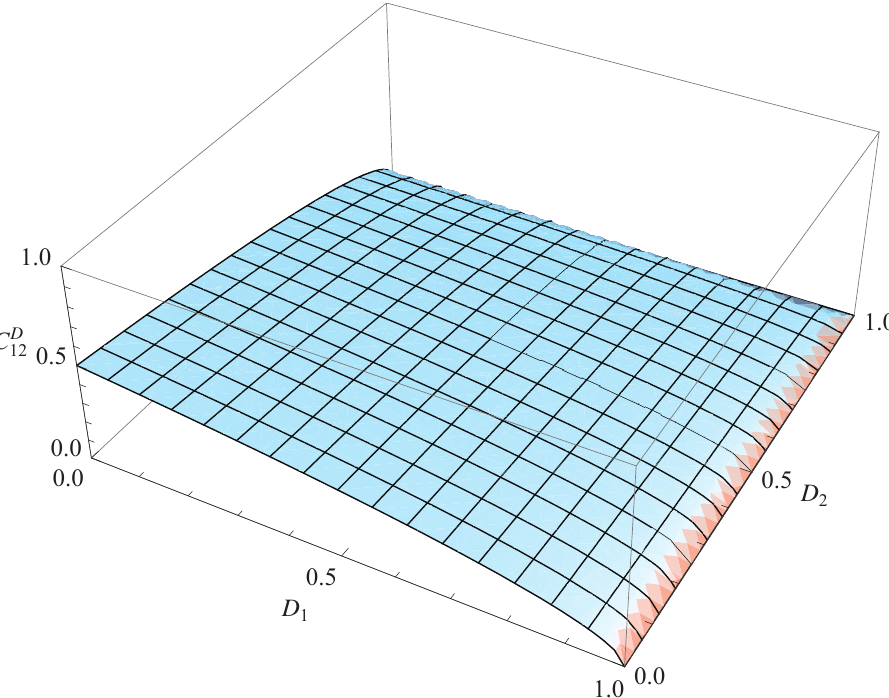}
\end{minipage} }
\subfigure[]
{ \begin{minipage}[b]{0.2\textwidth}
\includegraphics[width=1\textwidth]{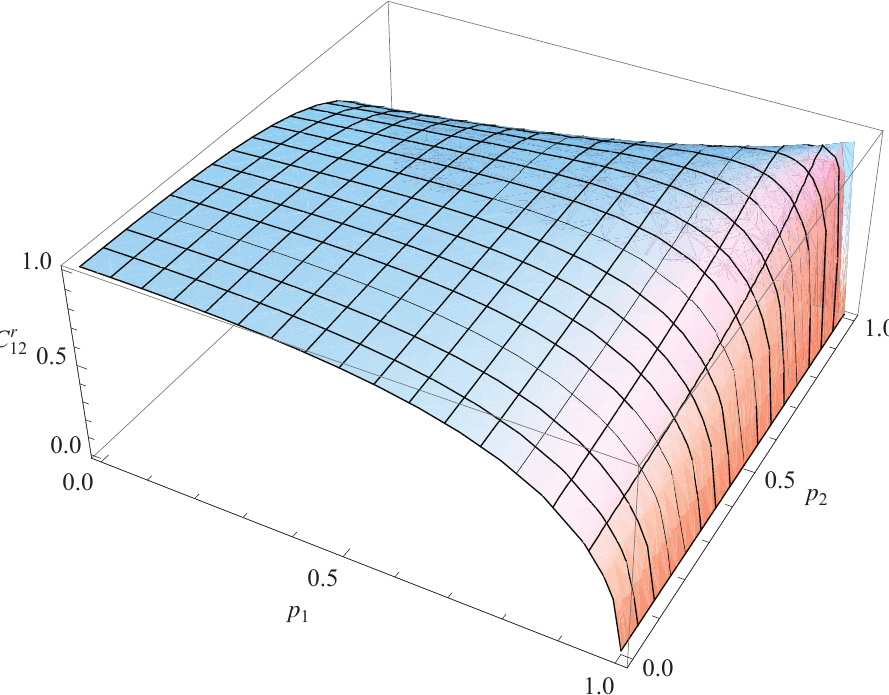}
\end{minipage} }
\subfigure[]
{ \begin{minipage}[b]{0.2\textwidth}
\includegraphics[width=1\textwidth]{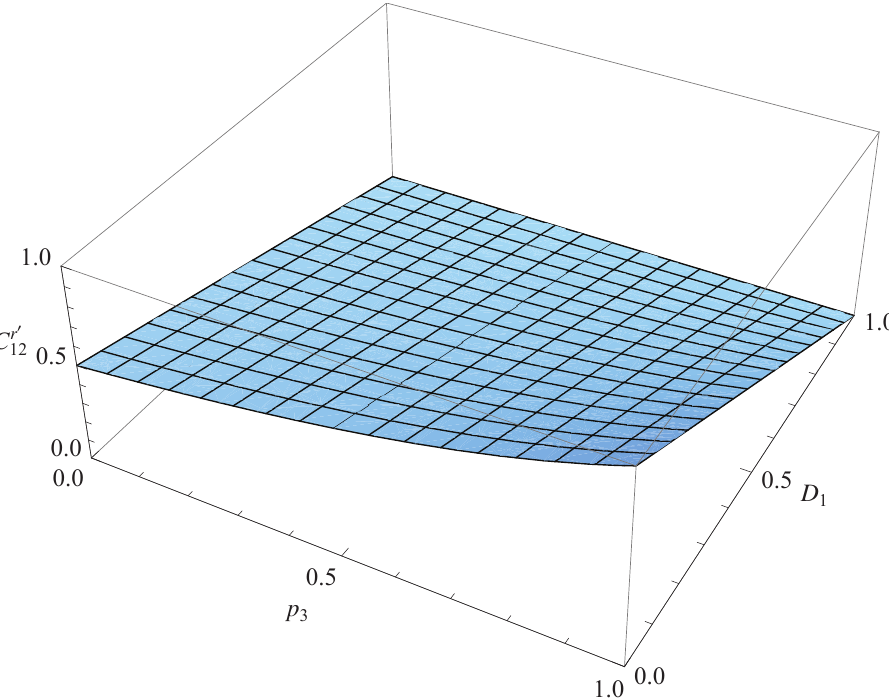}
 \end{minipage}
 }
 \caption{Comparison between distributed and fully local methods for taming decoherence under AD noise. In (a), the concurrence $C_{12}^{D}$ without the action of any measurement and under different decoherence strengths $D_1$ and $D_2$ is plotted; in (b), the concurrence $C_{12}^{r}$ under AD decoherence and
distributed reversed weak measurements is reported (for $D_1$ $=$ $D_2$ $=0.6$, $q_1$ $=$ $q_2$ $=0.99$ ); in (c), the
concurrence $C_{12}^{r'}$ under decoherence and fully local
method reversed weak measurements is plotted (for $D_1=D_2$,$q_3=0.99$ ).\quad \qquad \qquad  }
\label{fig:Fig2}
\end{figure}

An analogous analysis in the case of DP noise is summarised in Fig.~3. 
The explicit calculation of the concurrence through DP noise is rather convoluted and not very instructive, 
so we will just content ourselves with presenting numerical findings. 
It is interesting to point out that, in contrast with the AD case, here the fully local reversed weak measurement strategy outperforms the distributed one. The region with a large $p_3$ for protecting entanglement works better than that of a little $p_3$ in the fully local method. For more general noise models, the optimality of distributed versus fully local reversals will hence depend on the relative strength of the AD and DP sources of noise. Also, for more general qubit CP-maps, it is not trivial to determine which strategy provides in general a better performance.
\begin{figure}
\subfigure[]
{ \begin{minipage}[b]{0.2\textwidth}
\includegraphics[width=1\textwidth]{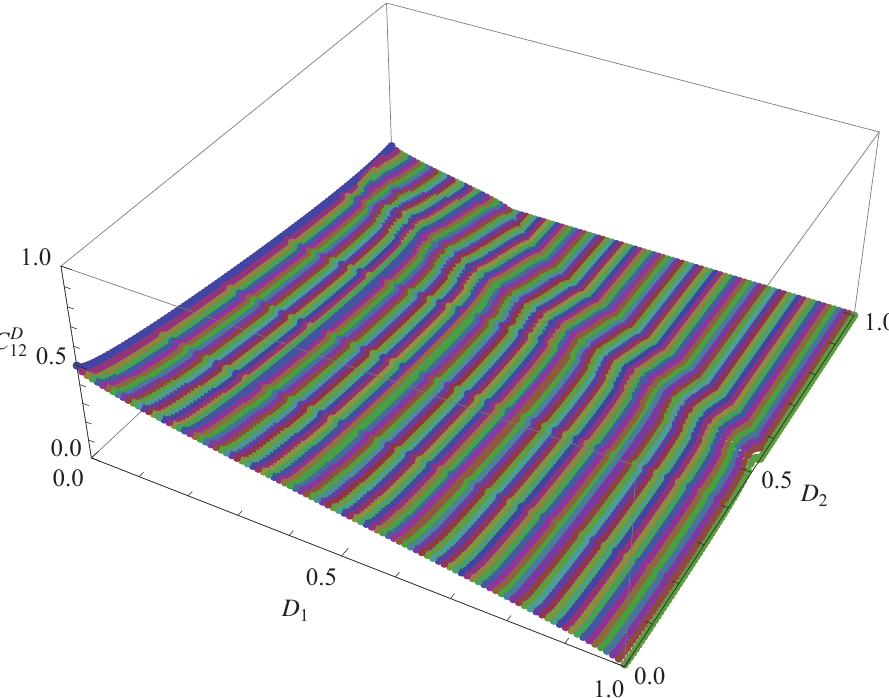}
\end{minipage} }
\subfigure[]
{ \begin{minipage}[b]{0.2\textwidth}
\includegraphics[width=1\textwidth]{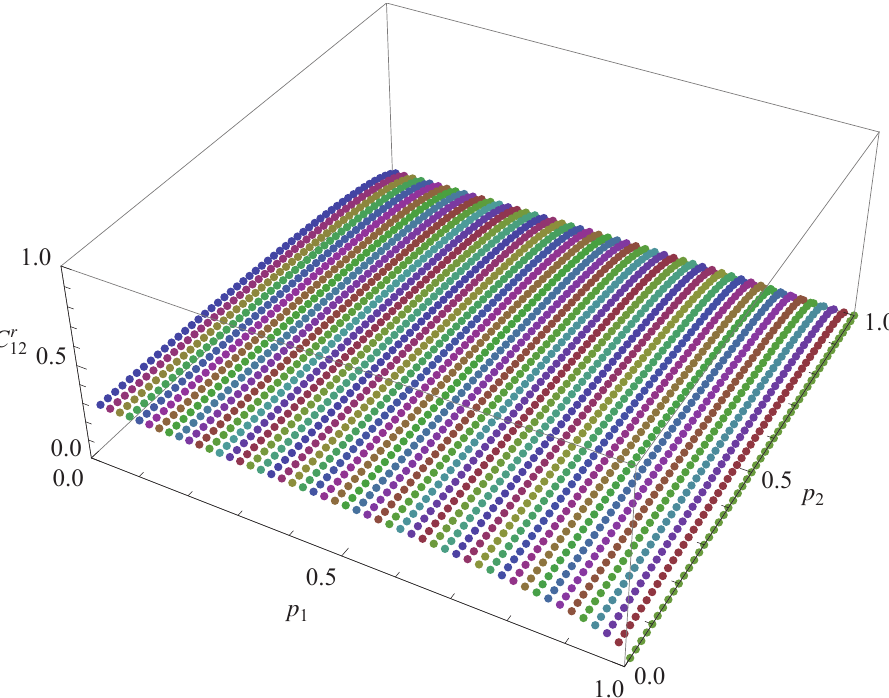}
\end{minipage} }
\subfigure[]
{ \begin{minipage}[b]{0.2\textwidth}
\includegraphics[width=1\textwidth]{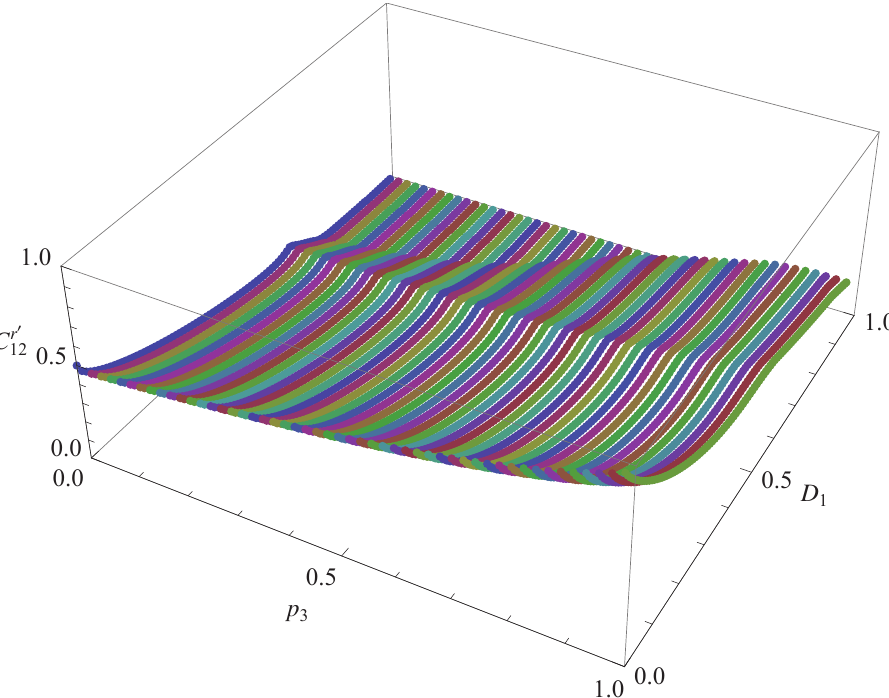}
\end{minipage} }
\caption{Comparison between distributed and fully local methods for taming decoherence under AD noise. In (a), the concurrence $C_{12}^{D}$ without the action of any measurement and under different decoherence strengths $D_1$ and $D_2$ is plotted; in (b), the concurrence $C_{12}^{r}$ under AD decoherence and
distributed reversed weak measurements is reported ($D_1=D_2=0.2$, while $q_1$ and $q_2$ take their optimal values); in (c), the
concurrence $C_{12}^{r'}$ under decoherence and fully local
method reversed weak measurements is plotted ($D_1=D_2$, while $q_3$ takes its optimal value).}
\label{fig:Fig3}
\end{figure}

Let us now consider the dependence of our analysis on the specific tripartite states considered so far. To this end, we have calculated $C_{12}^{D}$ and $C_{12}^{r}$ by varying the coefficients of a W-like state, reporting our results in Fig.~4. It was found that our method works well for different W-like states suffering from AD noise. As explicitly shown in Fig.~4b, an almost perfect EPR pair can be  generated with a certain probability in the case of $a_1=a_2$, which means that one can distill an EPR pair from a large family of W-like states with the assistance of the third party by using weak measurements and reversal. For the initial mixed state $\rho=\frac{0.1}{8}I_{8}+0.9 |GW\rangle \langle GW|$, 
where $|GW\rangle=$ $\frac{2}{5}|100\rangle$ $+\frac{2}{5}|010\rangle+\frac{\sqrt{17}}{5}|001\rangle$, our method still gets a pure entangled state of qubits 1 and 2.

\begin{figure}
\centering
\subfigure[]
{ \begin{minipage}[b]{0.2\textwidth}
\includegraphics[width=1\textwidth]{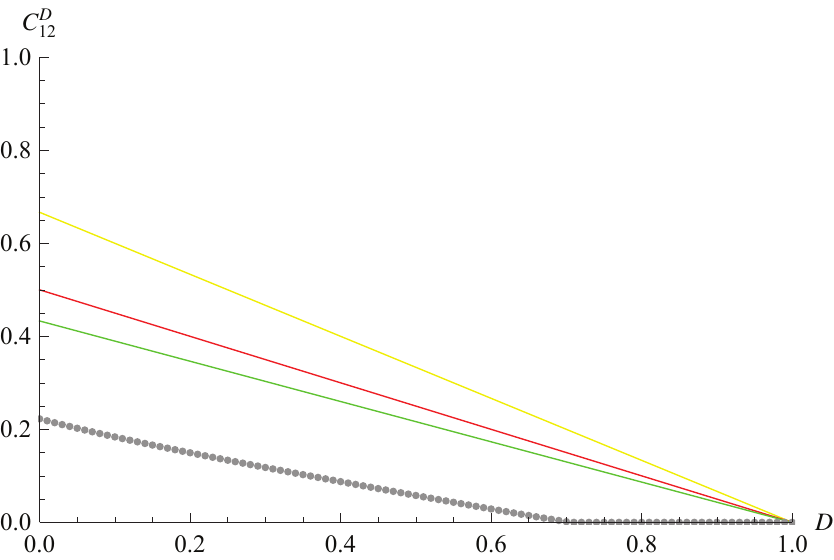}
\end{minipage} }
\subfigure[]
{ \begin{minipage}[b]{0.2\textwidth}
\includegraphics[width=1\textwidth]{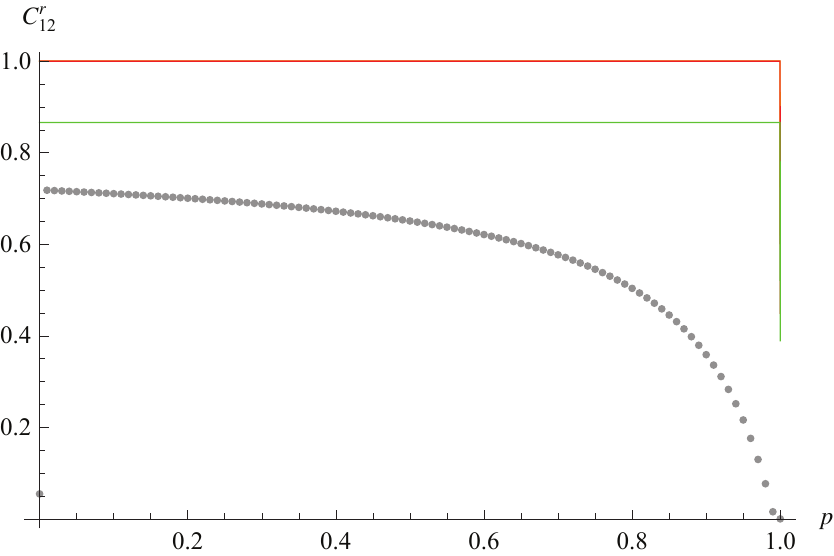}
\end{minipage} }
\caption{Distributed method for taming the AD decoherence of different initial W-like states. In (a), the concurrence $C_{12}^{D}$ of different W-like states under AD noise with no reversed weak measurement is plotted; in (b), the concurrence $C_{12}^{r}$ of different W-like states change under AD noise and distributed reversed weak measurements (for $p_1$ $ =$ $ p_2$,$q_1=q_2= 0.99$,$D_1$ $=$ $D_2$ $=$ $0.6$) is plotted; yellow color: the initial state is $\frac{1}{\sqrt{3}}|100\rangle+\frac{1}{\sqrt{3}}|010\rangle+\frac{1}{\sqrt{3}}|001\rangle$, red color: the initial state is $\frac{1}{2}|100\rangle+\frac{1}{\sqrt{2}}|010\rangle+\frac{1}{\sqrt{2}}|001\rangle$, green color: the initial state is $\frac{\sqrt{3}}{\sqrt{8}}|100\rangle+\frac{1}{\sqrt{8}}|010\rangle+\frac{1}{\sqrt{2}}|001\rangle$, grey color: the initial state is $\frac{0.1}{8}I_{8}+0.9 |GW\rangle \langle GW|$, where $|GW\rangle=$ $\frac{2}{5}|100\rangle$ $+\frac{2}{5}|010\rangle+\frac{\sqrt{17}}{5}|001\rangle$.}
\label{fig:Fig4}
\end{figure}

\begin{figure}
\includegraphics[scale=0.4]{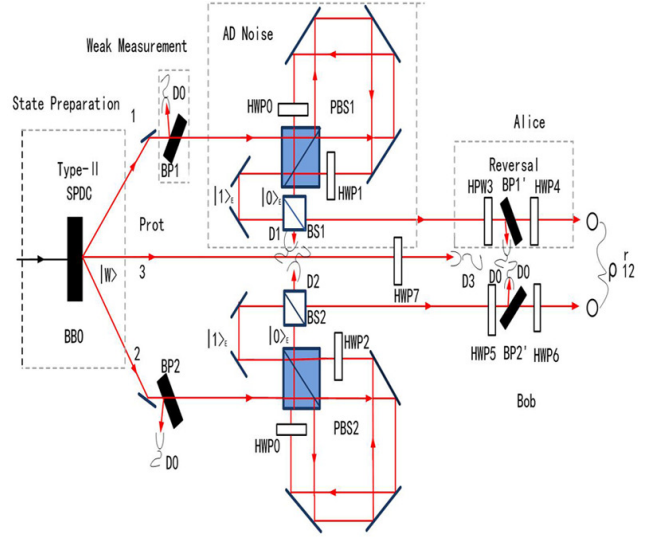}
\caption{Schematics of the proposed experimental set-up for suppressing AD quantum noise by weak measurements and distributed reversal. The polrization-entangled W state is prepared by type-II spontaneous parametric down-conversion (SPDC). A set of Brewster angle glass plates (BP) implement weak measurements reversals. The amplitude damping channel is simulated by Sagnac-type interferometers (SI).}
\label{fig:Fig 5}
\end{figure}

\section{Experimental implementation}\label{exp}

The AD noise, weak measurement, and its reversal can be implemented with linear optics, using the photon polarization as a qubit.
We propose an experimental set-up (shown in Fig. 5) for taming decoherence under AD noise by distributed weak measurement and reversal.

The three-photon polarization-entangled $W$-state $|W\rangle$ can be created by a second-order emission process of type-II spontaneous parametric
down-conversion (SPDC) [19], and can be written as
  \begin{equation}
\label{eq:12}
|W\rangle=\frac{|HHV>+|HVH>+|VHH>}{\sqrt{3}},
\end{equation}
where $H$ means horizontal polarization and $V$ means vertical  polarization.
The effect of amplitude damping can be simulated, and calibrated,
by Sagnac-type interferometers (SI) [20].
Before and after going through the SI, two photonic qubits 1 and 2 undergo a set of Brewster-angle glass plates (BP) and half-wave plates
(HWP) that implement weak measurement and its reversal in spatially separated locations, respectively. We assign the bit value 0 to the H polarization and 1 to the V polarization. Since $BP_i$ (here $i=1, 2$) only reflects the vertical polarization state $|1\rangle$ with a probability of reflection $p_i$,
the null event at single-photon detector $D_i$ in the reflected mode of $BP_i$ heralds that the single-photon found in the transmitted mode of $BP_i$ has been subjected to a weak measurement $M_w^i$. The reversing measurement $M_r^i$ is implemented with two half-wave plates for the bit-flip operations and $BP_i'$. Weak measurement and its reversal strength $p_i$ and $q_i$ can be varied by changing the number of $BP's$.

The decoherence of the AD noise effect shown in Eq.~(7) is realized using a SI with an additional beamsplitter (BS), which involves the tracing out of an environmental qubit. The displaced SI implements the coupling of the polarization qubit to the path qubit. The horizontal polarization $|0\rangle_s$ transmitting through the polarizing beam splitter $i (PBS_i)$ (here $i=1,2$) can be found only at the $|0\rangle_E$ output mode, corresponding to the state map
$|0\rangle_{S}|0\rangle_E\rightarrow|0\rangle_S|0\rangle_E$. On the other hand, the vertical polarization $|1\rangle_S$ entering the $PBS_i$ can be found both at $|0\rangle_E$ and $|1\rangle_E$ output modes depending on the angle $\theta_i$ of the $HWP_i$. The probability that the vertical polarization $|1\rangle_S$ at the input of the PBS ends up at the $|1\rangle_E$ output mode of the $PBS_i$ corresponds to the two result terms of the state map
$|1\rangle_S|1\rangle_E\rightarrow\sqrt{(1-D_i)}|1\rangle_S|0\rangle_E+\sqrt{D_i}|0\rangle_S|1\rangle_E$, with $ \sqrt{D_i}=\sin(2\theta_i)$. As we are interested only in the system's qubit, the environment's qubit is traced out by incoherently mixing $|1\rangle_E$ (horizontally polarized) and $|0\rangle_E$ (vertically polarized) at the BS. Note that the path-length difference between $|1\rangle_E$ and $|0\rangle_E$ is, accordingly, larger than the coherence length of the down-converted photons ($\thicksim140\mu m$).

Finally, the Hadamard gate H on photonic qubit 3 can be implemented by $HWP7$, given by
\begin{equation}
\label{eq:13}
U_{HWP}(\theta)=\left(
\begin{array}{cc}
\cos(2\theta) & \sin(2\theta)  \\
\sin(2\theta) & -\cos(2\theta) \\%
\end{array}%
\right),
\end{equation}
tilted at $\theta=\frac{\pi}{8}$. Thus, if four detectors $D0$ do not click, regardless of whether a click at the single-photon detector $D3$ occurs or not, the entanglement $\rho_{12}^r$ between photonic qubits 1 and 2 will have been obtained, with a positive concurrence $C_{12}^r$, independent from the outcome of the measurement at detector $D3$.

\section{Conclusions and outlook}

In summary, we explored a probabilistic approach to concentrate entanglement into two parties of W-like states based on weak measurements and their optimal reversal. Our main result is that
decoherence due to either amplitude damping and phase diffusion noise can be suppressed by exploiting quantum measurement reversal, in which a weak measurement and its optimal reversing measurement are introduced before and after decoherence channel, respectively. We also proposed an experimental implementation based on polarization degrees of freedom to test our findings. The approach investigated here provides an active way to tame decoherence and transfer entanglement, and we believe that this may shed some light on the problem of entanglement transformations as well.

It is natural to ask if, how and to what extent this tripartite-to-bipartite entanglement extraction approach can be applied to parties with more than three subsystems. This is an open question that we will discuss it in a later paper.

\acknowledgments
This work was supported by the National Natural Science Foundation of China under Grants 61144006 and 10901103, by the Foundation of China Scholarship Council and by the Educational Committee of the Hunan Province of China through the Overseas Famous Teachers Program.

\end{document}